\documentclass[uplatex, a4paper,12pt]{article} 
\usepackage[utf8]{inputenc}
\usepackage{amsmath}
\usepackage{color}
\usepackage{amssymb, amsfonts, amsthm, amscd}
\usepackage{bm}
\usepackage{float}
\usepackage{caption}
\usepackage[subrefformat=parens]{subcaption}
\usepackage{hyperref}  
\usepackage{graphicx}  

\usepackage{bookmark}            
\hypersetup{
  setpagesize=false,
  bookmarksnumbered=true,
  bookmarksopen=true,
  colorlinks=true,
  linkcolor=blue,
  citecolor=red,
}
\begin{document}

\thispagestyle{empty}
\vspace*{-15mm}

\begin{flushleft}
{\bf OUJ-FTC-15}\\

\end{flushleft}

{\bf }\

\vspace{15mm}

\begin{center}
{\Large\bf
On the Effects of Elementary and Bound State Fields on Vacuum Stability in $t\bar{t}$ Production at the LHC}

\baselineskip 18pt
\vspace{7mm}


Yoshiki Matsuoka

\vspace{5mm}

{\it Nature and Environment, Faculty of Liberal Arts, The Open University of Japan, Chiba 261-8586, Japan\\

}

\end{center}

\vspace{3cm}

\begin{flushleft} 
Email: machia1805@gmail.com  
\end{flushleft}

\vspace{10mm}
\begin{center}
\begin{minipage}{14cm}
\baselineskip 16pt
\noindent

\begin{abstract}
The recent report by the CMS Collaboration on the excess of top and anti-top pair production is examined with regard to the improvement of vacuum stability under two simple scenarios: one with the existence of toponium $(\eta_t)$, and the other with an additional elementary field $(\Psi)$. In both cases, the values of each coupling constant are restricted by the Multicritical Point Principle (MPP).

Two scenarios are considered. The first incorporates the effect of toponium $(\eta_t)$, inspired by the Bardeen-Hill-Lindner (BHL) framework, and the second embeds $\Psi$ into the Standard Model (SM) as a minimal model close to an inert doublet model. From these analyses, it is found that toponium cannot satisfy the MPP, whereas the additional elementary field can.

\end{abstract}

\end{minipage}
\end{center}

\vspace{0.5cm}
\newpage
\section{Introduction}
The Large Hadron Collider (LHC) continues to explore physics beyond the Standard Model (SM)\cite{higgs2,LHC1}. Recently, the CMS Collaboration has reported an excess of events in the invariant mass distribution of top-antitop pairs ($t\bar{t}$) near the threshold at $345$ GeV, based on the proton-proton collision data at $\sqrt{s} = 13\ \mathrm{TeV}$ with an integrated luminosity of $138\ \mathrm{fb}^{-1}$\cite{toponium}. This excess has a local statistical significance, and one of the most compelling interpretations is the formation of a pseudo-scalar bound state of toponium ($J^{\mathrm{PC}} = 0^{-+}$)\cite{toponium_1}. Another plausible interpretation is the existence of an additional unknown elementary field(pseudo-scalar and/or scalar).

To understand the recent CMS report from the viewpoint of improving vacuum stability, we examine two simple scenarios: one introducing a bound state (toponium) $\eta_t$, and the other introducing an additional elementary field $(\Psi)$. Thus, the purpose of this paper is to examine, in each scenario, whether the Higgs quartic coupling, after taking into account quantum corrections, remains positive at all energy scales $(\mu)$. We investigate whether these two scenarios can improve the stability of the vacuum. For this purpose, in this paper we determine the phenomenologically viable parameter region under the Multicritical Point Principle (MPP)\cite{MPPoriginal,MPPoriginal2,MPPoriginal3,kk1,mppeg1,kawai1,mppeg2,me1}.

In this study, even in the case including the bound state $\eta_t$, the renormalization group (RG) analysis is inevitable. To address this, we adopt the Bardeen-Hill-Lindner (BHL) framework\cite{topotopo,topotopo2}, in which a composite particle is treated as if it were elementary, but a different wavefunction renormalization condition, $Z_2(\mu=\Lambda)=0$, is imposed at a cutoff scale $\Lambda$ to maintain consistency with compositeness.

Furthermore, the model with the new elementary field $\Psi$ is closely related to the inert Higgs doublet model\cite{2hdm}. The differences from the inert Higgs doublet model are twofold. First, the coupling to the Higgs is restricted to the form $(H^\dagger H)(\Psi^\dagger \Psi)$. Second, among Yukawa couplings, only that to the top quark is present. In this paper, we investigate such a model.

The organization of this paper is as follows. In Section \ref{sec2}, we review the MPP and its application to the one-loop effective potential. In Section \ref{sec3}, we explain the BHL condition, present the model where toponium is embedded into the SM under the BHL condition, derive the renormalization group equations (RGEs), and analyze the parameter space consistent with MPP. In Section \ref{sec4}, we analyze in a similar way the model where an additional elementary field is incorporated into the SM. Section \ref{sec5} is devoted to summary and brief discussion. \ref{firstappendix} provides the one-loop RGEs of the SM with toponium, \ref{secondappendix} describes the one-loop RGEs of the SM with an additional elementary field.
\section{What is MPP?}\label{sec2}
In this section, we explain the MPP.  For $\mu$ as the renormalization scale and $h$ as the Higgs field, the MPP imposes the following conditions on the effective potential:
\begin{eqnarray}\label{mppcd}
V_{\mathrm{\mathrm{eff}}}(h=v_{\mathrm{EW}}, \mu_{\mathrm{EW}})&=&V_{\mathrm{eff}}(h=v_c, \mu_c)\simeq0,\\ \
\frac{dV_{\mathrm{eff}}(h=v_{\mathrm{EW}}, \mu_{\mathrm{EW}})}{d\mu}&=&\frac{dV_{\mathrm{eff}}(h=v_c, \mu_c)}{d\mu}\simeq0
\end{eqnarray}
where $V_{\mathrm{\mathrm{eff}}}$ is one-loop effective potential of the vacuum expectation value (VEV) of the Higgs scalar, and $\mu_{\mathrm{EW}}$ and $v_{\mathrm{EW}}$ are the Electroweak scale and VEV of $246$ GeV and $\mu_c\sim v_c>>\mu_{\mathrm{EW}}$ GeV is the MPP scale at a significantly high energy. The first equation expresses the equality of the effective potentials at different scales, which realizes the multi-criticality. It is important to observe that the two minima of the effective potential have the same value at different scales. In order to realize this, $V_{\mathrm{eff}}(\mu_c)$ must be approximately zero, since the Electroweak scale and the MPP scale $\mu_c$ are significantly different, without any relationship.

We impose the renormalization conditions as
\begin{eqnarray}
\left.\frac{dV_{\mathrm{eff}}}{dh}\right|_{h=v_{\text{EW}}} &=& 0,\\
\left.\frac{d^2V_{\mathrm{eff}}}{dh^2}\right|_{h=v_{\text{EW}}} &=& M^2_h,\\
\left.\frac{d^4V_{\mathrm{eff}}}{dh^4}\right|_{h=M_t} &=& 6\lambda(M_t)
\end{eqnarray}
where $M_t$ is the top quark mass and $M_h$ is the Higgs mass. 

The MPP has shown remarkable predictive power, notably in anticipating the top quark and Higgs masses\cite{MPPoriginal,MPPoriginal2}. This empirical success lends support to the idea that the effective potential is fine-tuned by unknown high scale dynamics to exhibit multi-criticality. These MPP conditions will be employed in the subsequent sections.

\section{The model of the SM with toponium}\label{sec3}
In this section, we study the model where toponium $\eta_t$ is added to the SM. In this case, in addition to the perturbative analysis, we incorporate the non-perturbative effects of toponium formation by imposing a specific boundary condition on the wavefunction renormalization constant, following the method proposed by Bardeen, Hill, and Lindner (BHL)\cite{topotopo,topotopo2}. Specifically, we introduce an effective four-fermion interaction between top quarks at a certain scale $\Lambda$, and take into account the resulting fermion loop effects. This induces the wavefunction renormalization of the bound state $\eta_t$.

In this framework, the bound state $\eta_t$ is introduced as an explicit degree of freedom and is described by assuming a Yukawa coupling with the top quark. We analyze the behavior of $Z_2(\mu)$ using the one-loop renormalization group equations, and impose the boundary condition $Z_2(\Lambda)=0$ at the cutoff scale $\Lambda$. This condition implies that $\eta_t$ can be effectively described as an elementary field, and provides the basis for constructing the low-energy effective theory. Additionally, we take VEV of $\eta_t$ to be zero. Furthermore, within this framework, the additional Yukawa interactions and the tree-level scalar potential involving these fields are given by:
\begin{eqnarray}
-\mathcal{L}_{\mathrm{Yukawa}}=y_{\eta_t}\bar{Q}_{3L}\tilde{\eta_t}t_R+(\mathrm{h.c.})
\end{eqnarray}
and
\begin{eqnarray}
V(H,\eta_t) &=& M_{\eta_t}^2(\eta_t^\dagger\eta_t)+\lambda_{\eta_t}(\eta_t^\dagger \eta_t)^2+\kappa_{1}(H^\dagger H)(\eta_t^\dagger \eta_t)
\end{eqnarray}
where $M_{\eta_t}$ denotes the mass of $\eta_t$ and the couplings $\kappa_1$ and $\lambda_{\eta_t}$ are set to vanish at a cutoff scale, $\kappa_1(\Lambda) = \lambda_{\eta_t}(\Lambda) = 0$\footnote{Here, the conditions $\kappa_1(\mu) \ge 0$ and $\lambda_{\eta_t}(\mu) \ge 0$ can always be maintained for all renormalization scales $\mu$ so that the $\eta_t$'s VEV remains zero.
}, while the Yukawa coupling is taken to be large, $y_{\eta_t}(\Lambda) = \infty\ (\geq \sqrt{4\pi})$. These represent the boundary conditions imposed on each coupling constant following the BHL approach. In the BHL approach the compositeness boundary condition is formulated as $y_{\eta_t}(\Lambda)\rightarrow \infty$. In practice, we replace the divergence by a strong, but finite, value and take $y_{\eta_t}(\Lambda) = \sqrt{4\pi}$. This choice is motivated by naive dimensional analysis and tree-level partial-wave unitarity for Yukawa interactions: once $y_{\eta_t}\sim \mathcal{O}(\sqrt{4\pi})$, the loop expansion parameter $\frac{y_{\eta_t}^2}{16\pi^2}$ ceases to be small and the coupling is effectively non-perturbative. Importantly, for RG flows starting from such large boundary values the low-energy trajectory is insensitive to the precise choice of the UV value, so using $\sqrt{4\pi}$ faithfully reproduces the $y_{\eta_t}(\Lambda)\rightarrow\infty$ limit while keeping the numerics stable.  

We examined whether the MPP conditions at the MPP scale near $\mu \simeq \mu_c$ are satisfied by the effective potential with respect to the BHL condition as follows:
\begin{eqnarray}\label{conditions}
\left.V_{\mathrm{eff}}\right|_{\mu=\mu_c}&=&0,\\
\label{conditions2}
\left.\frac{dV_{\mathrm{eff}}}{d\mu}\right|_{\mu=\mu_c}&=&0.
\end{eqnarray}
In our scenario, the one-loop effective potential of the SM in Landau gauge using $\mathrm{\overline{MS}}$ scheme reads\cite{higgs3}:
\begin{eqnarray}
V_{\mathrm{eff}}\left(h(\mu),\mu\right) &=& \frac{\lambda(\mu)}{4}h^4(\mu)\nonumber\\
&+&\frac{1}{64\pi^2}\left(3\lambda(\mu) h^2(\mu)\right)^2\left(\ln\frac{3\lambda(\mu) h^2(\mu)}{\mu^2}-\frac{3}{2}\right)\nonumber\\
&+&\frac{3}{64\pi^2}\left(\lambda(\mu) h^2(\mu)\right)^2\left(\ln\frac{\lambda(\mu) h^2(\mu)}{\mu^2}-\frac{3}{2}\right)\nonumber\\
&+&\frac{3\times2}{64\pi^2}\left(\frac{g_2(\mu)h(\mu)}{2}\right)^4\left(\ln\frac{\left(g_2(\mu)h(\mu)\right)^2}{4\mu^2}-\frac{5}{6}\right)\nonumber\\
&+&\frac{3}{64\pi^2}\left(\frac{\sqrt{g_2^2(\mu)+g_Y^2(\mu)}}2h(\mu)\right)^4\left(\ln\frac{\left(g_2^2(\mu)+g_Y^2(\mu)\right)h^2(\mu)}{4\mu^2}-\frac{5}{6}\right)\nonumber\\
&-&\frac{4\times3}{64\pi^2}\left(\frac{y_t(\mu)h(\mu)}{\sqrt{2}}\right)^4\left(\ln\frac{\left(y_t(\mu)h(\mu)\right)^2}{2\mu^2}-\frac{3}{2}\right)\nonumber\\
&+&\frac{4}{64\pi^2}\left(\frac{\kappa_1(\mu) h^2(\mu)}2\right)^2\left(\ln\frac{\kappa_1(\mu) h^2(\mu)}{2\mu^2}-\frac{3}{2}\right)
\end{eqnarray}
where $h(\mu),\lambda(\mu),y_t(\mu), g_Y(\mu), g_2(\mu)$ represent the Higgs field and SM coupling constants with each depending on $\mu$ which is the renormalization scale. We focus on examining the behavior near the MPP scale $\mu_c$. However, the mass term in the Higgs field is the Electroweak scale, and is sufficiently small.
Therefore, it is neglected in this effective potential. 

We examine the one-loop RGEs [\ref{firstappendix}]. Also, we simply put $h=\mu$ since the effect on the effective potential is negligibly small and $h(\mu)$ is the renormalized running field.\footnote{We do not distinguish between the bare field and the renormalized running field as its wave function renormalization is so small.}. 
As references, we use \cite{cern,PDG}. Using the top quark mass $M_t = 172.69 \pm 0.30\ \mathrm{GeV}$ and the strong coupling constant $\alpha_s(M_Z) = 0.1179 \pm 0.0010$\cite{PDG}, we numerically calculate whether the MPP conditions can be approximately satisfied under the BHL condition. Since the effects remain significant up to the region exceeding the toponium mass by several tens of GeV\cite{bound}, we choose $\Lambda = 370$--$400\ \mathrm{GeV}$. However, the impact of varying the cutoff scale  $\Lambda$ was negligible and the MPP conditions could not be satisfied in any case. Since the lower edge of the running window is set by the toponium threshold at approximately $345~\mathrm{GeV}$\cite{bound}, we assume that the effective four-fermion interaction responsible for binding top quarks first becomes operative at this scale. Accordingly, $\eta_t$-related terms are evolved only over the interval $345~\mathrm{GeV}\le \mu \le \Lambda$, with compositeness boundary conditions imposed at $\mu=\Lambda$ and matching to the SM at $\mu=345~\mathrm{GeV}$.

\begin{figure}[H]
 \begin{center}
 \includegraphics[width=100mm]{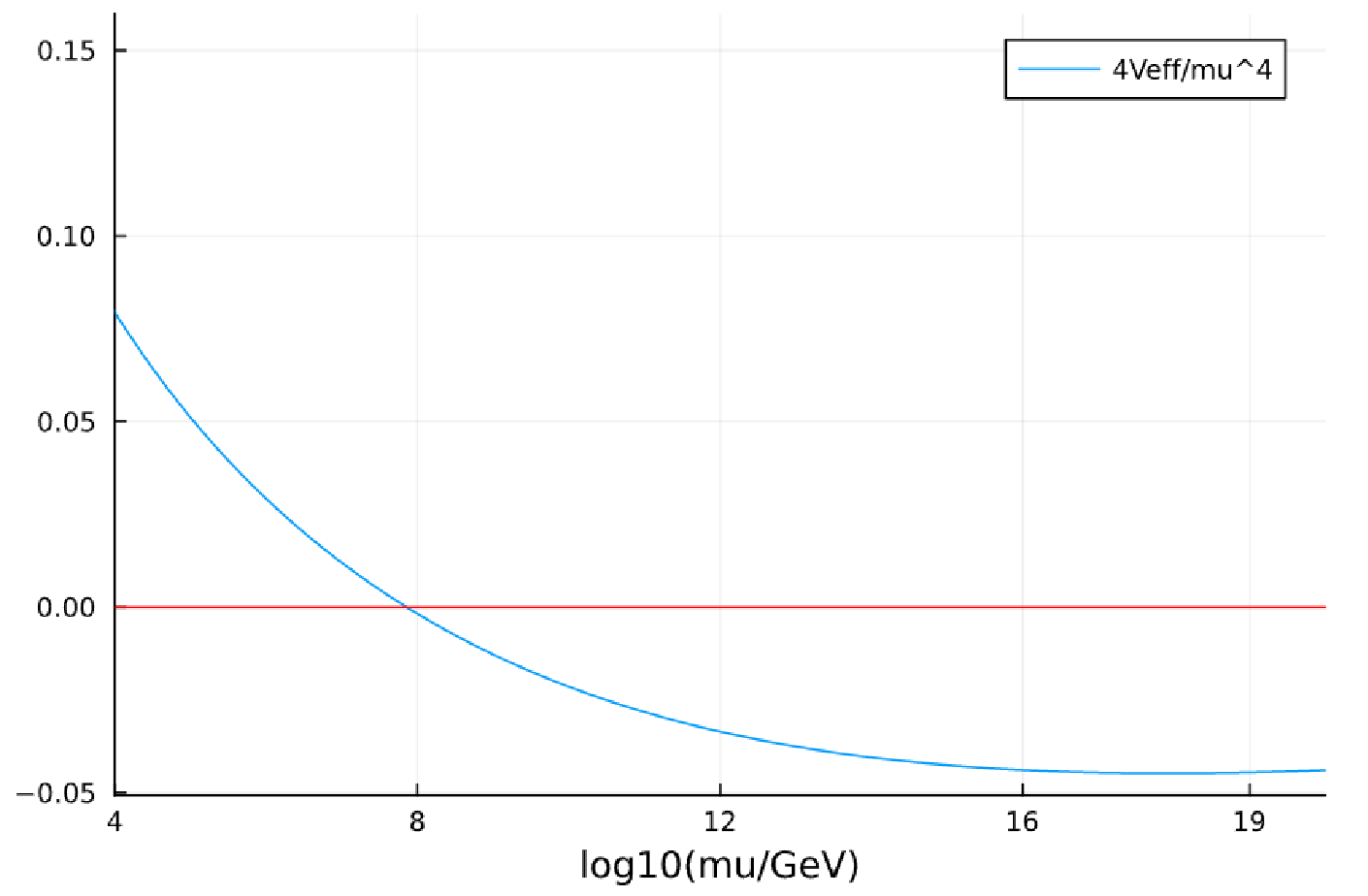}
 \caption{The $x$-axis and $y$-axis show $\log_{10}(\frac{\mu}{\mathrm{GeV}})$ and the value of each line. The blue line is $\frac{4V_{\mathrm{eff}}}{\mu^4}$ in the case of $M_t=172.69\ \mathrm{GeV},\ \alpha_s(M_Z)=0.1189$, and $\Lambda = 400\ \mathrm{GeV}$. The red line shows zero on the $y$-axis.}\label{fig1}
 \end{center}
\end{figure}

In Figure \ref{fig1}, for the case with $M_t=172.69\ \mathrm{GeV},\ \alpha_s(M_Z)=0.1189$, and $\Lambda = 400\ \mathrm{GeV}$, the behavior is plotted with the energy scale on the $x$-axis and the effective potential on the $y$-axis. It can be seen that the minimum value of the term for $\lambda(\mu)$ including quantum corrections goes below about $-0.04$. This indicates that the vacuum stability is deteriorating. This deterioration is assessed relative to the SM, where $\lambda(\mu)$ with quantum corrections included in the RGEs reaches approximately $-0.025$ at one loop and $-0.015$ at three loops\cite{higgs3,cern}. Such a substantially more negative behavior may be incompatible with the observed longevity of the Universe (i.e., a long-lived Electroweak vacuum) and is therefore problematic\cite{v1,v2}.
\section{The model of the SM with an additional elementary field}\label{sec4}
\begin{figure}[H]
 \begin{center}
 \includegraphics[width=100mm]{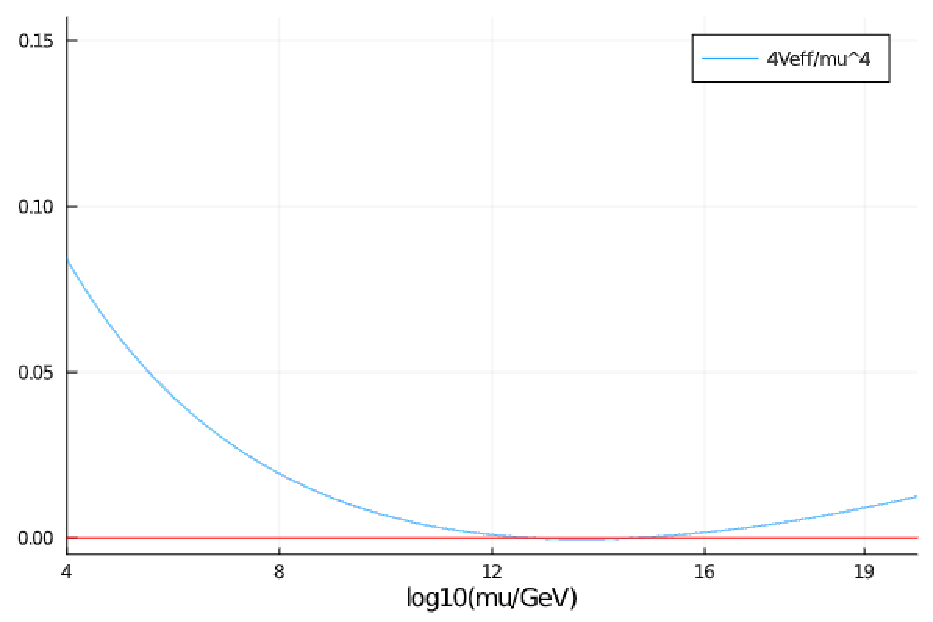}
  \caption{The $x$-axis and $y$-axis show $\log_{10}(\frac{\mu}{\mathrm{GeV}})$ and the value of each line. The blue line is $\frac{4V_{\mathrm{eff}}}{\mu^4}$ in the case of $M_t=172.69\ \mathrm{GeV},\ \alpha_s(M_Z)=0.1189, \ y_\Psi(M_t) = 0.4,$ and $\kappa_2(M_t) =0.244$. The red line shows zero on the $y$-axis. It can be seen that the MPP condition is satisfied around $\mu_c=10^{13.5}$ GeV.}\label{fig2}
 \end{center}
\end{figure}

In this Section , we consider a model in which an additional elementary field $\Psi$ is incorporated into the SM. This additional field is based on the inert Higgs doublet model, supplemented with a Yukawa coupling to the top quark. The additional Yukawa interactions and tree-level scalar potential are given by:
\begin{eqnarray}
-\mathcal{L}_{\mathrm{Yukawa}}=y_{\Psi}\bar{Q}_{3L}\tilde{\Psi}t_R+(\mathrm{h.c.})
\end{eqnarray}
and
\begin{eqnarray}
V(\Psi,H) &=& M_{\Psi}^2(\Psi^\dagger \Psi)+\lambda_\Psi(\Psi^\dagger \Psi)^2+\kappa_{2}(H^\dagger H)(\Psi^\dagger \Psi)
\end{eqnarray}
The one-loop effective potential in this case can be written as:
\begin{eqnarray}\label{ef}
V_{\mathrm{eff}}\left(h(\mu),\mu\right) &=& \frac{\lambda(\mu)}{4}h^4(\mu)\nonumber\\
&+&\frac{1}{64\pi^2}\left(3\lambda(\mu) h^2(\mu)\right)^2\left(\ln\frac{3\lambda(\mu) h^2(\mu)}{\mu^2}-\frac{3}{2}\right)\nonumber\\
&+&\frac{3}{64\pi^2}\left(\lambda(\mu) h^2(\mu)\right)^2\left(\ln\frac{\lambda(\mu) h^2(\mu)}{\mu^2}-\frac{3}{2}\right)\nonumber\\
&+&\frac{3\times2}{64\pi^2}\left(\frac{g_2(\mu)h(\mu)}{2}\right)^4\left(\ln\frac{\left(g_2(\mu)h(\mu)\right)^2}{4\mu^2}-\frac{5}{6}\right)\nonumber\\
&+&\frac{3}{64\pi^2}\left(\frac{\sqrt{g_2^2(\mu)+g_Y^2(\mu)}}2h(\mu)\right)^4\left(\ln\frac{\left(g_2^2(\mu)+g_Y^2(\mu)\right)h^2(\mu)}{4\mu^2}-\frac{5}{6}\right)\nonumber\\
&-&\frac{4\times3}{64\pi^2}\left(\frac{y_t(\mu)h(\mu)}{\sqrt{2}}\right)^4\left(\ln\frac{\left(y_t(\mu)h(\mu)\right)^2}{2\mu^2}-\frac{3}{2}\right)\nonumber\\
&+&\frac{4}{64\pi^2}\left(\frac{\kappa_2(\mu) h^2(\mu)}2\right)^2\left(\ln\frac{\kappa_2(\mu) h^2(\mu)}{2\mu^2}-\frac{3}{2}\right).
\end{eqnarray}
Assuming $\lambda(M_t) \sim \lambda_\Psi(M_t)$, taking the VEV of $\Psi$ to be zero\footnote{By setting the VEV of $\Psi$ to zero, one can avoid the constraint on a mixing angle between scalars at the LHC.} by imposing $\lambda_\Psi(\mu) \geq 0$ and $\kappa_2(\mu)>-2\sqrt{\lambda(\mu)\lambda_\Psi(\mu)}$, and using the CMS result\footnote{An analysis that augments the SM with only a scalar and a pseudo-scalar, in the sense of the CMS result, is not available. However, since the observed excess is dominated by the pseudo-scalar relative to the scalar, we adopt the constraints obtained for the pseudo-scalar only case. Using the $2\sigma$ lower edge of the exclusion bound on the Yukawa coupling for this pseudo-scalar, we determine the lower limit of the Yukawa coupling.}\cite{toponium},
\begin{eqnarray}
|y_\Psi(M_t)| \geq 0.4.
\end{eqnarray}
We find that the MPP conditions can be satisfied when
\begin{eqnarray}
\kappa_2(M_t) \geq 0.244,\ \mu_c \leq 10^{13.5}\ \mathrm{GeV}
\end{eqnarray}
where we use Eq. (\ref{ef}) and the one-loop RGEs [\ref{secondappendix}]. The behavior of the effective potential is illustrated in Figure \ref{fig2}, where it can be clearly observed that the MPP condition is satisfied in the vicinity of $\mu_c = 10^{13.5}$ GeV.

\section{Summary and Discussion}\label{sec5}
In this paper, motivated by the excess of $t\bar{t}$ events near the threshold region reported by the LHC, we have investigated the behavior of the effective potential when toponium and, separately, an additional field are incorporated into the SM. In the toponium case, we constructed an effective model that incorporates the non-perturbative effects associated with its formation, whereas with an additional field we constructed an inert Higgs doublet–like model. 

To describe the non-perturbative effects associated with toponium formation, we adopted the BHL framework\cite{topotopo,topotopo2}, which imposes a boundary condition on the wavefunction renormalization of toponium. This condition reflects the compositeness of the toponium state and naturally leads to the emergence of a new Yukawa coupling. 

Within this framework, we introduced toponium as if it were an elementary field and applied the MPP as a high scale constraint. The vacuum becomes less stable. Furthermore, we carried out a similar MPP analysis for the case where an additional field is incorporated into the SM. In this setup, we found that the MPP conditions-vacuum stability- can be satisfied. 

These results indicate that the vacuum stability can be improved when an additional field is incorporated into the SM. However, when toponium is introduced into the SM, the vacuum stability is instead worsened in both the one-loop effective potential and RGEs. This is in comparison with the case in the SM, where $\lambda(\mu)$ with quantum corrections included in the RGEs reaches approximately $-0.025$ at one loop and about $-0.015$ at three loops\cite{higgs3,cern}. This result may be incompatible with the observational fact that the Universe has persisted to the present day\cite{v1,v2}. Therefore, instead of considering toponium alone, it is important to consider New Physics, or a framework in which toponium is combined with New Physics. Further investigations including higher-loop corrections are essential, and it is also necessary to experimentally evaluate the validity of the observed excess in $t\bar{t}$ events. As an immediate extension, it will be important to carry out a similar study within the framework of the Two-Higgs-Doublet Models (2HDM)\cite{DM2,Obs4,Obs5}. In any case, the results reported by the LHC are highly intriguing and may open a path toward New Physics.

\section*{Acknowledgments}
We thank Noriaki Aibara, Gi-Chol CHO, So Katagiri, Shiro Komata, and Akio Sugamoto for many helpful comments. Especially, I would like to take this opportunity to extend my deepest appreciation to Gi-Chol CHO and Akio Sugamoto for their generous advice and the fruitful discussions we shared.

\appendix
\renewcommand{\thesection}{Appendix \Alph{section}}
\section{One-loop Renormalization Group Equations including a new field and toponium}\label{firstappendix}
The one-loop RGEs are
\begin{eqnarray}
\frac{dg_Y}{dt}&=&\frac{g_Y^3}{16\pi^2}(\frac{41}6+\frac{1}{6}\delta),\\ \frac{dg_2}{dt}&=&\frac{g_2^3}{16\pi^2}(-\frac{19}6+\frac{1}{6}\delta),\\ \frac{dg_3}{dt}&=&\frac{g_3^3}{16\pi^2}(-7), \\ 
\frac{dy_t}{dt}&=&\frac{y_t}{16\pi^2}\Big(\frac{9}2y_t^2+\frac{3}2y_{\eta_t}^2-\frac{17}{12}g_Y^2-\frac{9}4g_2^2-8g_3^2\Big),\\
\frac{dy_{\eta_t}}{dt}&=&\frac{y_{\eta_t}}{16\pi^2}\Big(\frac{9}2y_{\eta_t}^2+\frac{3}2y_t^2-\frac{17}{12}g_Y^2-\frac{9}4g_2^2-8g_3^2\Big),\\
\frac{d\lambda}{dt} &=& \frac{1}{16\pi^2}\Big(\lambda\left(24\lambda-3g_Y^2-9g_2^2+12y_t^2\right)\nonumber \\
&+&2\kappa_{1}^2+\frac{3}{8}g_Y^4+\frac{3}{4}g_Y^2g_2^2+\frac{9}{8}g^4_2-6y_t^4\Big),\\
\frac{d\kappa_{1}}{dt} &=& \frac{1}{16\pi^2}\Big(\kappa_{1}\left(4\kappa_{1}+12\lambda+6\lambda_{\eta_t}-3g_Y^2-9g_2^2+6y_t^2+6y_{\eta_t}^2\right)\nonumber\ \\
&+&\frac{3}{4}g_Y^4-\frac{3}{2}g_Y^2g_2^2+\frac{9}{4}g^4_2-6y_t^2y_{\eta_t}^2\Big),\ \\
\frac{d\lambda_{\eta_t}}{dt} &=& \frac{1}{16\pi^2}\Big(\lambda_{\eta_t}\left(24\lambda_{\eta_t}-3g_Y^2-9g_2^2+12y_{\eta_t}^2\right)\nonumber\ \\
&+&2\kappa_{1}^2+\frac{3}{8}g_Y^4+\frac{3}{4}g_Y^2g_2^2+\frac{9}{8}g^4_2-6y_{\eta_t}^4\Big)
\end{eqnarray}
where $t=\ln\mu$. $\mu$ is the renormalization scale. $\delta=1$ when $345\ \mathrm{GeV}\le\mu\le\Lambda$ and $\delta=0$ otherwise.
\section{One-loop Renormalization Group Equations including a new field}\label{secondappendix}
The one-loop RGEs are
\begin{eqnarray}
\frac{dg_Y}{dt}&=&\frac{g_Y^3}{16\pi^2}(7),\\ \frac{dg_2}{dt}&=&\frac{g_2^3}{16\pi^2}(-3),\\ \frac{dg_3}{dt}&=&\frac{g_3^3}{16\pi^2}(-7), \\ 
\frac{dy_t}{dt}&=&\frac{y_t}{16\pi^2}\Big(\frac{9}2y_t^2+\frac{3}2y_\Psi^2-\frac{17}{12}g_Y^2-\frac{9}4g_2^2-8g_3^2\Big),\\
\frac{dy_\Psi}{dt}&=&\frac{y_\Psi}{16\pi^2}\Big(\frac{9}2y_\Psi^2+\frac{3}2y_t^2-\frac{17}{12}g_Y^2-\frac{9}4g_2^2-8g_3^2\Big),\\
\frac{d\lambda}{dt} &=& \frac{1}{16\pi^2}\Big(\lambda\left(24\lambda-3g_Y^2-9g_2^2+12y_t^2\right)\nonumber \\
&+&2\kappa_{2}^2+\frac{3}{8}g_Y^4+\frac{3}{4}g_Y^2g_2^2+\frac{9}{8}g^4_2-6y_t^4\Big),\\
\frac{d\kappa_{2}}{dt} &=& \frac{1}{16\pi^2}\Big(\kappa_{2}\left(4\kappa_{2}+12\lambda+6\lambda_\Psi-3g_Y^2-9g_2^2+6y_t^2+6y_\Psi^2\right)\nonumber\ \\
&+&\frac{3}{4}g_Y^4-\frac{3}{2}g_Y^2g_2^2+\frac{9}{4}g^4_2-6y_t^2y_\Psi^2\Big),\ \\
\frac{d\lambda_\Psi}{dt} &=& \frac{1}{16\pi^2}\Big(\lambda_\Psi\left(24\lambda_\Psi-3g_Y^2-9g_2^2+12y_\Psi^2\right)\nonumber\ \\
&+&2\kappa_{2}^2+\frac{3}{8}g_Y^4+\frac{3}{4}g_Y^2g_2^2+\frac{9}{8}g^4_2-6y_\Psi^4\Big).
\end{eqnarray}


\end{document}